\documentclass[conference]{IEEEtran}
\usepackage{cite}
\usepackage{graphicx}
\usepackage{dsfont}
\usepackage{subfigure}
\usepackage{amsmath}
\usepackage{mathtools}
\usepackage{amsmath, graphicx, cite, color, epsfig}
\usepackage{ragged2e}
\usepackage{amssymb}

\ifCLASSINFOpdf
\else
\fi

\begin{document}

%
\title{Optimizing Pilot Overhead for Ultra-Reliable Short-Packet Transmission}

\author{\IEEEauthorblockN{Mohammadreza Mousaei}
\IEEEauthorblockA{Department of Electrical and Computer Engineering\\
University of Illinois at Chicago, 
Chicago, Illinois\\
Email: mmousa3@uic.edu
\and
\IEEEauthorblockN{Besma Smida, {\it Senior Member,  IEEE}}
\IEEEauthorblockA{Department of Electrical and Computer Engineering\\
University of Illinois at Chicago, 
Chicago, Illinois\\
Email: smida@uic.edu}}}


%


\maketitle

\begin{abstract}
In this paper we optimize the pilot overhead for ultra-reliable short-packet transmission and investigate the dependence of this overhead on packet size and error probability.  In particular, we  consider a  point-to-point communication in which one sensor sends messages to a central node, or base-station, over AWGN with Rayleigh fading channel. We formalize the optimization in terms of approximate achievable rates at a given block length, pilot length, and error probability. This leads to more accurate pilot overhead optimization. Simulation results show that it is important to take into account the packet size and the error probability when optimizing the pilot overhead. 
\end{abstract}


%
\IEEEpeerreviewmaketitle

\section{Introduction}
Wireless network research has traditionally focused on increasing the information rate to meet the demand generated by human-operated mobiles \cite{Mohsen1, Vahid2, Mohsen2, Parsa2}. However, all sorts of autonomous machines ``things" with communication capabilities will soon need to be connected as well. The data transmitted to and from autonomous machines is very different from the data to and from human-operated mobile devices \cite{Saeid3}. The autonomous machines exchange a massive number of short data bursts at moderate data rates\cite{Saeid5,Vahid1,Saeid4} but with stringent reliability requirements.  These data bursts may result from industrial automation, wireless coordination among vehicles, smart grid control functions, or health-monitoring activities \cite{5G, Mohsen4, Saeid1, Parsa1}.  The central challenge with these new wireless services is that current wireless systems are not properly designed to support high-reliable short-packet transmission. \\
The goal of this research is to increase packet efficiency by optimizing the pilot overhead for short packet transmission. In practical communication schemes, one send packets, each of which has bits dedicated to control overhead, pilots for channel estimation, and data payload. In short-packet communications, low packet efficiency is a concern: a packet typically carries less than 40\% to 50\% of actual data and the relative proportions allocated to different portions must be carefully optimized \cite{short}. 
\subsection{Related works -- Pilot overhead optimization}
The optimization of pilot overhead, predicated on the maximization of the {\it ergodic channel capacity}, has been largely studied in the literature \cite{cavers1991, medard2000, Mohsen5,balter2001, mmse3, mmse1, mmse2, opt_tr2,ce_bem3, dong2004,  furrer2007, Vahid3}.  
In more common systems, where the pilot symbol power is fixed,  the optimization is over the number of pilot symbols. In that case, some explicit results have been established in both low and high power regimes. Numerical solutions are derived for  general power levels.  
By maximizing a tight lower-bound of the average
channel capacity, a closed-form solution for the average rate of pilot symbol in block-fading \cite{mmse3} and in continuous fading with rectangular Doppler  spectrum is derived in \cite{mmse2}.  
More recently, the optimization of the pilot overhead in a unified continuous and block-fading model is investigated in \cite{lozano2008,jindal2010}, and the dependence of the optimum overhead on various system parameters of interest (e.g., fading rate, signal-to-noise ratio) is quantified. 
Optimization posed in prior works is predicated on the maximization of the {\it ergodic channel capacity}.  They all assume that the packet error probability can be made arbitrarily small by choosing the packet length sufficiently large. This optimization, based on large block-length, is unsuitable for short-packet transmission. Indeed, we need a new analysis of the achievable rate to assess the performance of short-packet communication \cite{poly}. Unfortunately, the exact value of achievable rate is unknown even for channel models that are much simpler to analyze than the one encountered in wireless communications \cite{short}. Polyanskiy and {\it al.} recently provided a unified approach to obtain tight bounds on achievable rate by providing lower bound that coincides with an upper bound in \cite{poly}. They showed that for various channels, the data-rate varies with packet sizes, desired error probabilities, and channel dispersions. 

\subsection{Contributions}
The key departure from prior work on pilot overhead optimization is that we (a) use achievable-rate tight-bound expressions that are more accurate for short-packet transmission, (b) derive the minimum mean square error for continuous fading as function of the packet size, and (c) investigate the dependence of the pilot overhead on various system parameters, e.g. packet size, error probability, fading rate and signal-to-noise ratio (SNR). 
%

%

\section{Preliminaries}
We consider a  point-to-point communication, in which  one sensor wishes to send messages to a central node, or base-station. The sensor sent training symbols known to the base-station, enabling the base-station to estimate the channel gain.

\subsection{Channel Model}
In this system, we consider a Rayleigh fading channel and additive white Gaussian noise (AWGN).  Pilot symbols are periodically inserted in every packet. Let each packet contain $n$ symbols. The transmission is divided into 
two phases. The training phase includes $n_t$ symbols and the data transmission phase includes $n-n_t$ symbols. We define parameter $\alpha = n_t/n$\footnote{ $\alpha$ should be greater than $\alpha_{min} = 1/n$.}.
Under this model,  the input-output relationship of $i^{th}$ received symbol is given by:
\begin{equation} y(i) = \sqrt{\rho}x(i)h(i) + w(i), \; i = 0,1,\ldots,n-1\end{equation}
where $x(i)$ is the $i^{th}$ symbol, $y(i)$ is the corresponding received symbol, $w(i)$  is  AWGN with zero-mean. Without
loss of generality, we normalize the Rayleigh fading channel ($|\mathbf{h}|^2=1$), and we assume $x(i)$ and $w(i)$ have unit mean square. Thus, $\rho$ is the SNR at the receiver.



%


\subsection{Channel Estimation} 
The $n_t$ training symbols are used  to estimate $h(i)$ for all $i$ in the data transmission phase.   We  first evaluate  the minimum  channel estimation error of the channel vector  $\mathbf{h}:=[h(0), \ldots, h(n)]^T$, as function  of $n_t$, which is needed to subsequently derive the approximate achievable rate. Let $\widetilde{\mathbf{h}}=\mathbf{h}-\widehat{\mathbf{h}}$ denote the mismatch between the true channel vector $\mathbf{h}$ and its estimate $\widehat{\mathbf{h}} :=[\widehat{h}(0), \ldots, \widehat{h}(n)]^T$. In this paper, we use minimum mean square error (MMSE) estimator with two different fading models:

1) \textit{Block Fading:} The block-fading model applies to a channel in which several adjacent symbols (referred as a block or packet) are affected by the same fading value and the fading values in different blocks are independent and identically distributed\cite{Saeid2}. For example, this model is applicable to wearable health sensors which are transferring short-packet communication (such as body temperature and heart bit rate) to smartphones. Communication environment in such system changes in a low speed so that the channel gain, albeit random, varies so slowly with time that it can be assumed as constant along a block. Using this fading model we can derive MMSE as $\sigma^2_{\mathbf{\widetilde{h}}}$ as  \cite{mmse3}: 
\begin{equation}
\label{5} \sigma^2_{\mathbf{\widetilde{h}}} = \dfrac{1}{1 + \alpha n \rho}. 
\end{equation}

2) \textit{Continuous Fading:} For wireless communication the continuous fading channel model is more realistic. Indeed, the channel is continuously changing, so the actual channel will deviate progressively from the channel estimate obtained at the training time. The channel estimation error for continuous fading channels is caused by noise as well as the temporal variation of the channel \cite{deltah}. We can hence model our channel as a block fading channel with an additional noise due to the  temporal variation of the channel.  
Assuming this model we can derive  $\sigma^2_{\mathbf{\widetilde{h}}}$ 
\begin{eqnarray}
\sigma^2_{\mathbf{\widetilde{h}}} = \dfrac{1}{1 + \alpha n \rho} + \sigma^2_{Doppler},
\end{eqnarray}
where  the additional channel estimation error  $\sigma^2_{Doppler}$ is derived, in the appendix, for Rayleigh fading as 
\begin{eqnarray}
\sigma^2_{Doppler} &=& 2\left(\dfrac{\pi\alpha n\rho f_D}{1 + \alpha n \rho}\right)^2\left( n - \dfrac{\alpha n}{2} \right)^2.
\end{eqnarray}
where $f_D$ is the Doppler frequency normalized to the data rate. 


\subsection{Ergodic Capacity in Finite Blocklength Regime}
Consider a source which is modeled as a random variable equi-probably taking values in the set $\{1, \dots, M\}$. The channel is a noisy communication medium which takes an input in some alphabet $\mathcal{A}$ and output a symbol in alphabet $\mathcal{B}$. An encoder maps messages ($\{1, \dots, M\}$) into length $n$ sequences of channel input symbols $\mathcal{A}^n$("codewords"). Therefore, the encoder is a function $f: \{1, \dots, M\} \rightarrow \mathcal{A}^n$. A decoder that produces an estimate of original signal by observing n-sequence of channel outputs is a function $g: \mathcal{B}^n \rightarrow \{1, \dots, M\}$. The goal of communication is to find an encoder-decoder pair (code) which is capable of communicating messages with some fixed probability of error $\epsilon$. Such code is called ($n, M, \epsilon$)-code. 
Polyanskiy and {\it al.} proved  in~\cite{poly, poly2} that given a fixed block-length $n$, probability of error $\epsilon$ and fading channel with SNR=$\rho$ and perfect CSI, the maximum number of messages 
$
M^*(n,\epsilon, \rho) := \max\{ M, \exists (n, M, \epsilon)-\mbox{code} \},
$
can be tightly approximated by
\begin{equation}
\label{sym}
\log_2(M^{*}(n, \epsilon, \rho)) \approx n\mathds{C}(\rho) - \sqrt{n\mathds{V}(\rho)}Q^{-1}(\epsilon),
\end{equation}
where  \begin{eqnarray}
   \label{eq: 8}
 \mathds{C}(\rho) = \log_2(e) e^{1/\rho} E_1(\dfrac{1}{\rho}), 
	\end{eqnarray}
and $E_1(x) = \int_1^\infty t^{-1}e^{-xt} dt$ is the exponential integral. The channel dispersion $\mathds{V}(\rho)$ can be derived as \cite{poly2}:

\begin{eqnarray}
\label{1001}
\mathds{V}(\rho) &=& \mbox{Var}\left[C(\rho |\mathbf{h}|^2)\right] \nonumber\\
&&+ \dfrac{\log^2(e)}{2} \left(1 - \mbox{E}^2_{|\mathbf{h}|^2}\left[\dfrac{1}{1 + \rho |\mathbf{h}|^2}\right]\right), \nonumber \\
\hspace{-2cm}
\end{eqnarray}
 where $\mbox{Var}[.]$  and $\mbox{E}[.]$ are variance and expectation over distribution of $|\mathbf{h}|^2$, and $Q^{-1}$ is functional inverse of $Q$-function. The ratio $R(n, \epsilon,\rho) := \dfrac{\log_2 M(n, \epsilon, \rho)}{n}$ is known as the rate. The maximum achievable rate can be tightly approximated by \cite{poly}:
\begin{equation}
\label{1} R^{*} (n, \epsilon,\rho) := \dfrac{1}{n} \log_2(M^{*}(n, \epsilon,\rho)) \approx \mathds{C}(\rho) - \sqrt{ \dfrac{\mathds{V}(\rho)}{n} } Q^{-1} (\epsilon).
\end{equation} 

\section{Pilot-assisted Detection for Short-Packet Transmission}
In this section, we provide an approximation of the achievable rate of a point-to-point communication($R_{\textit{Tr}}^*(n, \epsilon, \rho)$) when training symbols and MMSE estimator are used to extract CSI at the receiver. Contrary to the assumption in Section. II-C, here $\mathbf{h}$ is not known to the receiver but estimated using pilot symbols. 
 During the data transmission phase, after MMSE estimation channel can be rewritten as:
\begin{equation}
\label{chMMSE}
y(i) = \sqrt{\rho}x(i)\widehat{h}(i) + \sqrt{\rho}x(i)\widetilde{h}(i) + w(i),
\end{equation}
 where  the channel state information, $\widehat{h}(i)$, is perfectly known at receiver. The main problem here is the fact that the noise  $\sqrt{\rho}x(i)\widetilde{h}(i) + w(i)$  includes the channel estimation error. So the noise is not necessarily independent from the transmitted signal nor Gaussian.  First, we assume MMSE estimator, then $\widetilde{h}(i)$ and $\widehat{h}(i)$ are orthogonal. Then we consider Gaussian noise, following the same approach used in ~\cite{mmse3}. Using those assumptions,  the channel defined in Eq. (\ref{chMMSE}) became similar to the channel introduced in Section II-C, with SNR 
\begin{eqnarray}
\rho_{\textit{eff}} &=&  \dfrac{\rho(1 - \sigma^2_{\mathbf{\widetilde{h}}})}{1 + \rho\sigma^2_{\mathbf{\widetilde{h}}}}. 
\end{eqnarray} 

Finally, we took into account the number of symbols dedicated to pilot, $n_t$ symbols don't carry data, to approximate  the achievable rate as: 
\begin{equation}
\label{32}
R_{\textit{Tr}}^*(n, \epsilon, \rho_{\textit{eff}}) \approx (1 - \alpha)\mathds{C}(\rho_{\textit{eff}}) - Q^{-1}(\epsilon)\sqrt{\dfrac{(1 - \alpha)\mathds{V}(\rho_{\textit{eff}})}{n}}.
\end{equation}

\begin{figure}
\centering
	\includegraphics[width=9.5cm]{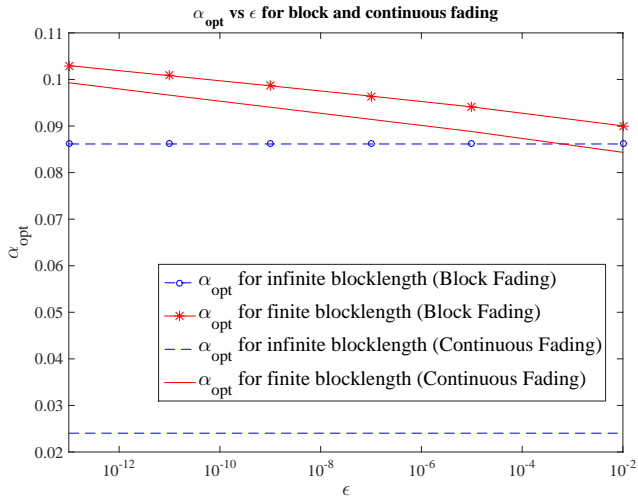}
	\vspace{-0.4cm} \caption{Optimal pilot overhead for infinite and finite blocklength in block and continuous fading model vs. $\epsilon$ with $n = 30$ and SNR = 15dB.}	
	\label{both_eps}
\end{figure}

\begin{figure}
\centering
	\includegraphics[width=9.5cm]{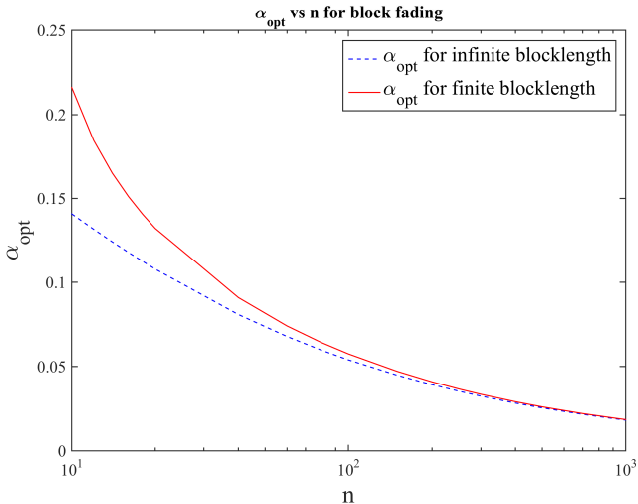}
	\vspace{-0.4cm}\caption{Optimal pilot overhead for infinite and finite blocklength in block fading model vs. Blocklength with SNR = 8dB and $\epsilon$ = 1e-9.}
	\label{block_n}
\end{figure}

\begin{figure}
\centering
	\includegraphics[width=9.5cm]{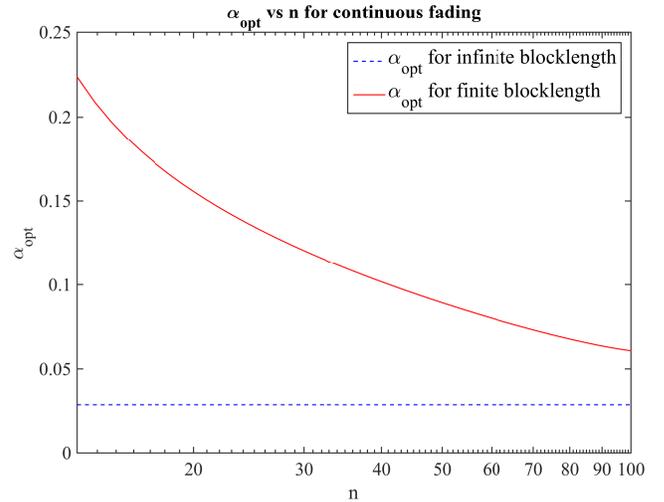}
	\vspace{-0.4cm}\caption{Optimal pilot overhead for infinite and finite blocklength in continuous fading model vs. Blocklength with SNR = 23dB, $\epsilon$ = 1e-9 and $f_D$ = 0.02.}	
	\label{cont_n}
\end{figure}

\section{Numerical Results}
\begin{figure}
\centering
	\includegraphics[width=9.5cm]{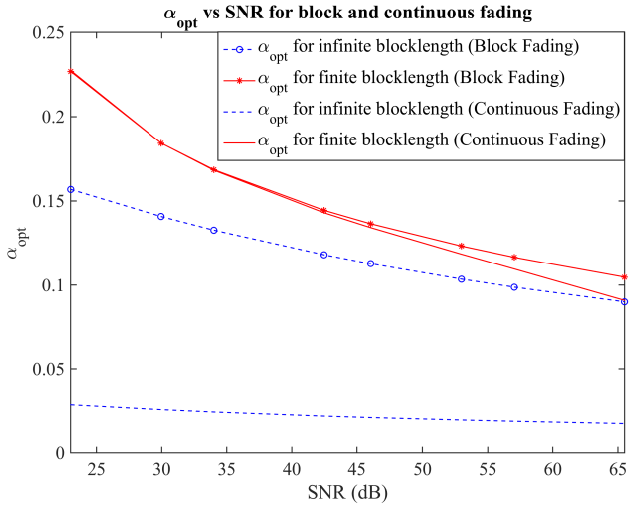}
	\vspace{-0.4cm}\caption{Optimal pilot overhead for infinite and finite blocklength in block and continuous fading model vs. SNR with $n = 40$ and $\epsilon$ = 1e-9.}	
	\label{both_SNR}
\end{figure}

\begin{figure}
\centering
	\includegraphics[width=9.5cm]{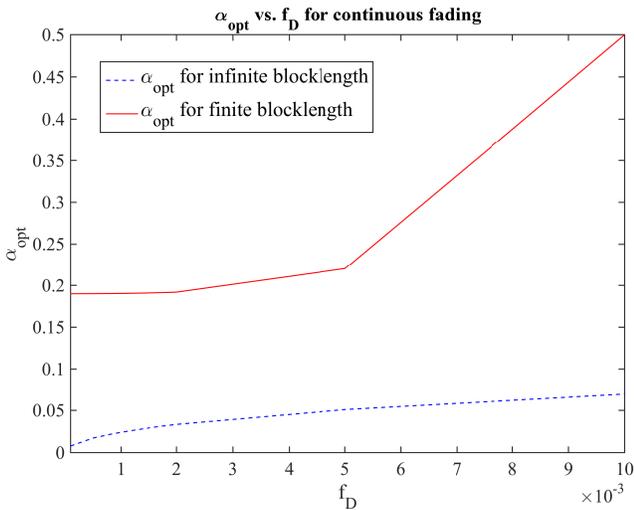}
	\vspace{-0.4cm}\caption{Optimal pilot overhead for infinite and finite blocklength in continuous fading model vs. $f_D$ with $n = 10$, SNR = 16dB and $\epsilon$ = 1e-9.}	
	\label{cont_fd}
\end{figure}

\begin{figure}
\centering
	\includegraphics[width=9.5cm]{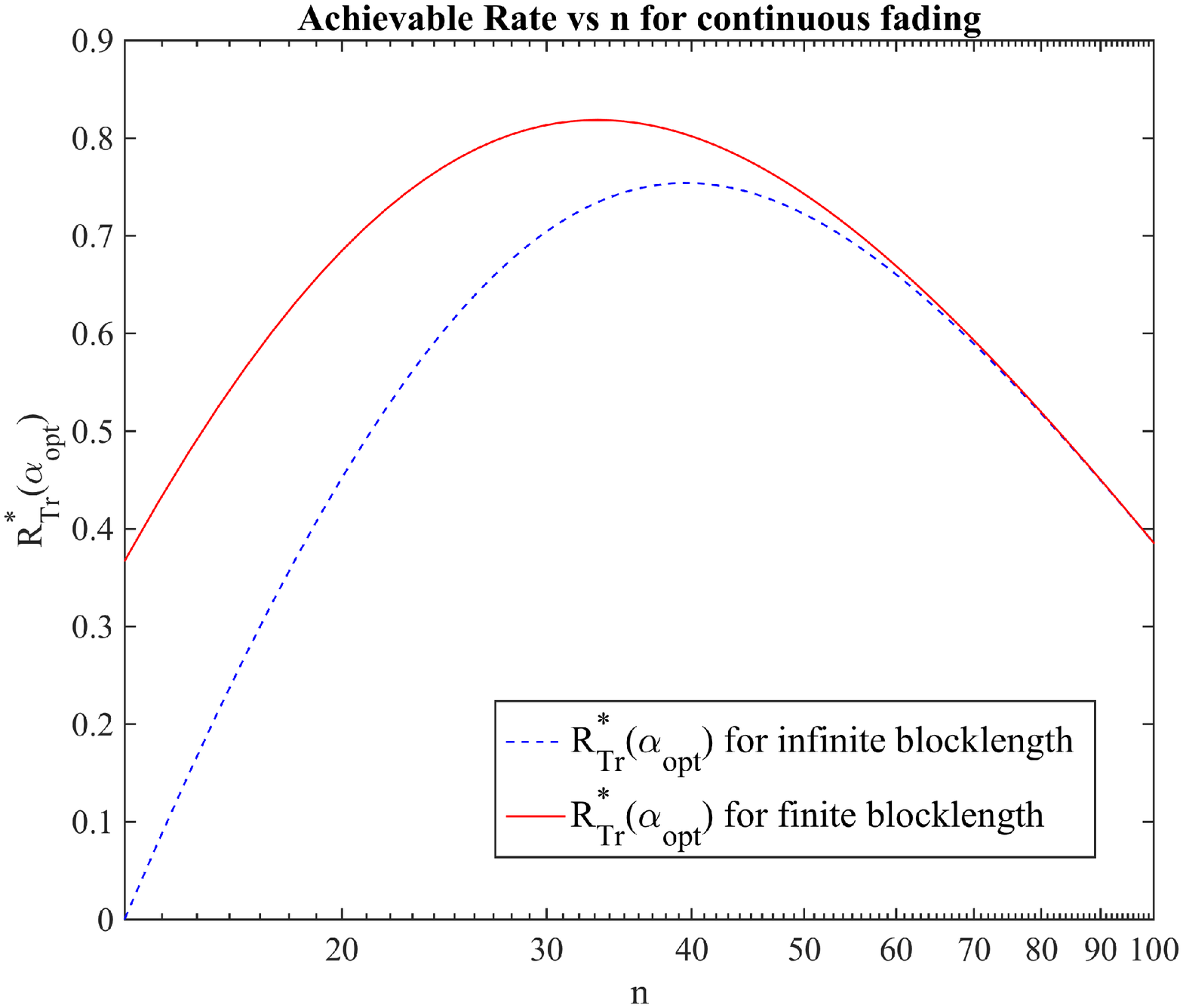}
	\vspace{-0.4cm}\caption{Acheivable rate using infinite and finite blocklength $\alpha_{opt}$ in continuous fading model vs. $n$ with SNR = 20dB, $\epsilon$ = 1e-12 and $f_D$ = 0.02.}	
	\label{Ropt_cont}
\end{figure}

\begin{figure}
\centering
	\includegraphics[width=9.5cm]{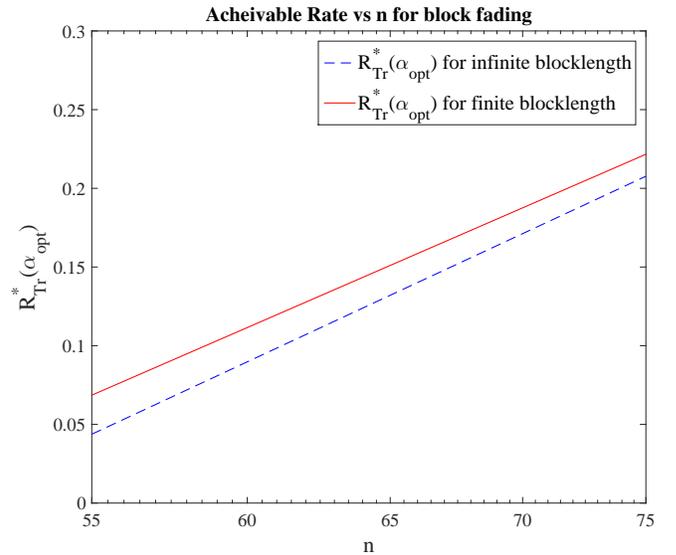}
	\vspace{-0.4cm}\caption{Achievable rate using infinite and finite blocklength $\alpha_{opt}$ in block fading model vs. $n$ with SNR = 7dB, $\epsilon$ = 1e-9 and $f_D$ = 0.02.}	
	\label{Ropt_block}
\end{figure}

In this section, we numerically evaluate the optimal pilot overhead for ultra-reliable short-packet transmission. These numerical results are derived by solving the derivative of Eq. (\ref{32}) w.r.t $\alpha$ equal to zero.  For comparison purpose we also optimize the pilot overhead using the ergodic capacity  \cite{lozano2008}.   
These comparisons are shown in Fig. \ref{both_eps}-\ref{Ropt_block}.  Our simulation results prove that our optimization approach will result in increase of around 10\% in achievable rate.  We performed simulations for a broad range of variables such as:

\begin{itemize}
\item \textbf{Probability of Error:} We first compare the optimal pilot overhead for different error probabilities. The difference between our optimal pilot overhead values and those derived using ergodic capacity increases with decreasing probability of error. Thus, it is  very important to use the new formulation for ultra-reliable communication systems. We have similar results for both block and continuous fading (Fig. \ref{both_eps}).

\item \textbf{Blocklength:}  We numerically evaluated the optimal pilot overhead for different blocklength. We considered both block and continuous fading. Note that when we use ergodic capacity optimization, the blocklength is assumed infinite but the variance of the channel estimation error varies with $n$ for block fading.   
As shown in Fig. \ref{block_n}, the difference is higher in small blocklength. This suggests that our approach is more adequate to short packet transmission. Moreover, we can see in Fig. \ref{cont_n} that the pilot overhead increases with small blocklength when we need more and more pilot symbols to compensate the channel estimation mismatch.

\item \textbf{SNR} and  \textbf{Normalized Doppler Frequency:}  The simulations, illustrated in Fig. \ref{both_SNR} and Fig. \ref{cont_fd},  show that the optimal pilot overhead decreases with SNR and increases with $f_D$, as expected. In addition,  the difference between our optimal pilot overhead values and those derived using ergodic capacity  is greater at low SNR and high normalized Doppler frequency. 

\item \textbf{Optimal Rate:} We evaluated the rate at the optimum values of $\alpha$ evaluated in this paper and optimum alpha for infinite blocklength. As illustrated in Fig. \ref{Ropt_block} our optimization approach will result in increase of roughly 10\% in rate with block fading. Fig. \ref{Ropt_cont} shows even more significant increase in rate with continuous fading. Simulations show that our approach always results in a higher rate. Fig. \ref{Ropt_cont} also shows that due to the channel estimation mismatch in continuous fading, increasing blocklength after a certain blocklength ($n = 29$ in this case) results in decreasing rate. 

\end{itemize}

\section{Conclusion}
The goal of this research is to increase the packet efficiency by optimizing the pilot overhead for ultra-reliable short packet transmission.  We  considered a  point-to-point communication in which  one sensor sends messages to a central node, or base-station, over additive white Gaussian noise with Rayleigh fading channel.   We formalized the optimization in terms of approximate achievable rates as function of block length, pilot length, and error probability. 
Simulation results proved that it is very important to take into account the packet size and the error probability when optimizing the pilot overhead. 
\section*{Acknowledgment}
Funding for this research was partially supported by NSF under Award Number 1620794.

\section*{Appendix}
The mobile transmit $n_t$ training symbols known to the mobile and base-station, enabling the base-station to estimate the channel gain.
The MMSE channel gain estimator can be derived as \cite{mmse3}
\begin{eqnarray}
\widehat{\textbf{h}} &=& \sqrt{\rho} (1 + \rho|\textbf{x}_t|^2)^{-1} \textbf{x}^*_t \textbf{y}_t \nonumber\\
&=& \dfrac{1}{\sqrt{\rho}} \left( \dfrac{1}{\rho} + |\textbf{x}_t|^2 \right)^{-1} \textbf{x}^*_t \textbf{y}_t
\end{eqnarray}
where  $\textbf{x}_{t}$, $\textbf{y}_{t}$ are input and output training symbol vectors. Note that 
$\textbf{y}_t = \sqrt{\rho} \textbf{x}_{t} \textbf{h} + \textbf{w}_t + \sqrt{\rho} \textbf{x}_t\Delta\textbf{h}_t,$ where $\textbf{y}_t = [y(1), y(2), \dots, y(\alpha n)]$, $\textbf{x}_t = [x(1), x(2), \dots, x(\alpha n)]$, $\textbf{w}_t = [w(1), w(2), \dots, w(\alpha n)]$ are output, input and noise vectors respectively. Also $\Delta\textbf{h}_t = [\Delta h(1), \Delta h(2), \dots, \Delta h(\alpha n)]$ is the channel mismatch due to the temporal variation of the channel.  
Since $|\textbf{x}_t|^2 = \alpha n$, we get
\begin{eqnarray}
\widehat{\textbf{h}} &=& \sqrt{\dfrac{1}{\rho}} \left(\dfrac{\rho}{1 + \alpha n \rho} \right) \left[\sqrt{\rho} \alpha n \textbf{h} + \right. \nonumber\\
&& \left. \textbf{w} \textbf{x}^*_t + \sqrt{\rho} \Delta\textbf{h}_t  \textbf{x}^*_t\textbf{x}_t \right] \nonumber\\
&=& \left(\dfrac{\alpha n \rho}{1 + \alpha n \rho} \right) \textbf{h} + \sqrt{\dfrac{1}{\rho}} \left( \dfrac{\rho}{1 + \alpha n \rho} \right) \textbf{w}\textbf{x}^*_t \nonumber\\
&& + \left(\dfrac{\alpha n \rho}{1 + \alpha n \rho} \right) \Delta\textbf{h}_t.
\end{eqnarray}
Hence the channel estimation error  $\widetilde{\textbf{h}} = \textbf{h} - \widehat{\textbf{h}} $ can be derived as: 
\begin{eqnarray}
\widetilde{\textbf{h}} &=& \textbf{h} - \widehat{\textbf{h}} \nonumber\\
&=& \dfrac{1}{1+\alpha n \rho} \textbf{h} - \sqrt{\dfrac{1}{\rho}} \left( \dfrac{\rho}{1 + \alpha n \rho} \right) \textbf{w}_t\textbf{x}^*_t \nonumber\\
&& - \left(\dfrac{\alpha n \rho}{1 + \alpha n \rho} \right) \Delta\textbf{h}_t
\end{eqnarray}
Finally, we derive the mean square error of the MMSE channel estimation error as 
\begin{eqnarray}
\sigma^2_{\widetilde{\textbf{h}}} = \dfrac{1}{1+\alpha n \rho} + \sigma^2_{{Doppler}}
\end{eqnarray}
where we derived $\sigma^2_{{Doppler}}$  using the mathematical derivation proposed in \cite{deltah} for maximum likelihood (ML) channel estimator in Rayleigh fading channel
\begin{eqnarray}
\sigma^2_{Doppler} &=& 2\left(\dfrac{\pi\alpha n\rho f_D}{1 + \alpha n \rho}\right)^2\left( n - \dfrac{\alpha n}{2} \right)^2,
\end{eqnarray}
where $f_D$ is the Doppler frequency normalized to the symbol rate ($R_{symbol} = 1/T_s$) given by $\dfrac{T_s v f_c}{c}$, $T_s$ is the symbol period, $v$ is the mobile velocity, $f_c$ is carrier frequency and $c$ is the speed of electromagnetic wave.

\bibliographystyle{IEEEtran}
\bibliography{refs_mohammad}{}

\end{document}